\documentclass[reprint,aps,pra,superscriptaddress]{revtex4-2}

\usepackage{graphicx}
\usepackage{dcolumn}
\usepackage{bm}
\usepackage[colorlinks=true,allcolors=blue]{hyperref}%
\usepackage[utf8]{inputenc}
\usepackage[english]{babel}

\usepackage{hyperref} 

\usepackage{booktabs}
\usepackage{colortbl}
\usepackage{multirow}

\usepackage{amsmath, amsthm, amscd, amssymb} 

\usepackage[mathlines]{lineno}

\newcommand{\delete}[1]{}

\begin{document}


\title{Precision calibration of the Duffing oscillator with phase control}

\author{Marc T. Cuairan}
\thanks{equal contribution}
\affiliation{ICFO Institut de Ciencies Fotoniques, Mediterranean Technology Park, 08860 Castelldefels (Barcelona), Spain}
\affiliation{Nanophotonic Systems Laboratory, Department of Mechanical and Process Engineering, ETH Zurich, 8092 Zurich, Switzerland}
\affiliation{Quantum Center, ETH Zurich, 8083 Zurich, Switzerland}
\author{Jan Gieseler}
\thanks{equal contribution}
\email{jan.gieseler@pm.me}
\affiliation{ICFO Institut de Ciencies Fotoniques, Mediterranean Technology Park, 08860 Castelldefels (Barcelona), Spain}
\author{Nadine Meyer}
\affiliation{ICFO Institut de Ciencies Fotoniques, Mediterranean Technology Park, 08860 Castelldefels (Barcelona), Spain}
\affiliation{Nanophotonic Systems Laboratory, Department of Mechanical and Process Engineering, ETH Zurich, 8092 Zurich, Switzerland}
\affiliation{Quantum Center, ETH Zurich, 8083 Zurich, Switzerland}
\email{nmeyer@ethz.ch}
\author{Romain Quidant}
\affiliation{ICFO Institut de Ciencies Fotoniques, Mediterranean Technology Park, 08860 Castelldefels (Barcelona), Spain}
\affiliation{Nanophotonic Systems Laboratory, Department of Mechanical and Process Engineering, ETH Zurich, 8092 Zurich, Switzerland}
\affiliation{Quantum Center, ETH Zurich, 8083 Zurich, Switzerland}
\affiliation{ICREA-Instituci\'{o} Catalana de Recerca i Estudis Avan\c{c}ats, 08010 Barcelona, Spain}

\date{\today}

\begin{abstract}
The Duffing oscillator is a nonlinear extension of the ubiquitous harmonic oscillator and as such plays an outstanding role in science and technology. 
Experimentally, the system parameters are determined by a measurement of its response to an external excitation.
When changing the amplitude or frequency of the external excitation, a sudden jump in the response function reveals the nonlinear dynamics prominently.
However, this bistability leaves part of the full response function unobserved, which limits the precise measurement of the system parameters.
Here, we exploit the often unknown fact that the response of a Duffing oscillator with nonlinear damping is a unique function of its phase.
By actively stabilizing the oscillator's phase we map out the full response function. This phase control allows us to precisely determine the system parameters. 
Our results are particularly important for characterizing nanoscale resonators, where nonlinear effects are observed readily and which hold great promise for next generation of ultrasensitive force and mass measurements.
We demonstrate our approach experimentally with an optically levitated particle in high vacuum. 
\end{abstract}


\maketitle

\section{Introduction}
\label{sec:introduction}

Nonlinear dynamics are abundant in nature and enable a wide variety of applications, such as precision measurements \cite{Clealand_NoisePrecision,Papariello_forcesensing_2016,Atalaya_2010,Zhao_nonlin_mass_Sensing_2018,buks_mass_2006, villanueva_surpassing_2013}, signal amplification \cite{chowdhury_weak_2020,siddiqi_rf-driven_2004,almog_signal_2007,Rodriguez_enhancingNoise_2020, ricci_optically_2017,chowdhury_phasestochastic_2017}, studies of the classical to quantum transition \cite{Katz_nonlinQuantumSignatures,arosh_quantumvanderpol_2021} and chaos \cite{Karabalin_2009,thompson2002nonlinear,moon1980experiments,testa1982evidence}. 
Most nonlinear oscillators are very well described by the lowest order nonlinear coefficients. For spatially symmetric oscillators the two dominant nonlinear contributions are the damping coefficient and amplitude squared dependent spring constant, respectively. The latter is also known as the Duffing or ${\chi}^3$ nonlinearity.\\
Precision measurements require to determine all system parameters precisely and accurately. This is done through calibration measurements and model fitting. Generally, the richer the dataset employed in model fitting the higher is the accuracy, and the bigger the dataset is the higher is the precision of the parameters.\\
Here, we show that with active phase stabilization of a driven nonlinear nanomechanical oscillator we are able to probe the otherwise unstable branch of the oscillator's response function \cite{MillerJ_PhaseControl}.
In contrast to previous work, we demonstrate phase control both for parametric and direct driving. Thereby, we create a calibration dataset that is significantly richer than what has been used in previous attempts to characterize the nonlinear coefficients \cite{flajsmanova_using_2020,Setter_nonlin_mecsqueezing_2019,eichler_nonlinear_2011,gieseler_nonlin_mode_coupling_2014,gieseler_thermal_2013,ricci_optically_2017,aldridge_noise-enabled_2005}. 
This dataset allows us reduce the error significantly and to show that parametric driving introduces a large systematic error.

The nanomechanical resonator is a levitated nanoparticle in our experiments. Levitated nanoparticles standout among nanomechanical systems through their high level of isolation from the environment and the in-situ tunability of the system parameters, without the need for re-fabrication. Hence, their are perfectly suited to study nonlinear effects \cite{ricci_optically_2017}, force and inertial sensing  \cite{gieseler_thermal_2013,Ranjit_zeptonewton_2016,Hebestreit_staticforce,Hempston_force,Monteiro_force_acceleration_sensing_2020} and to test the fundamentals laws of quantum mechanics \cite{tebbenjohanns_quantum_2021,magrini_real-time_2021}.

\section{Experimental setup}
\label{sec:exp_setup}
\begin{figure}[h]
	\centering
	\includegraphics[width=0.75\linewidth]{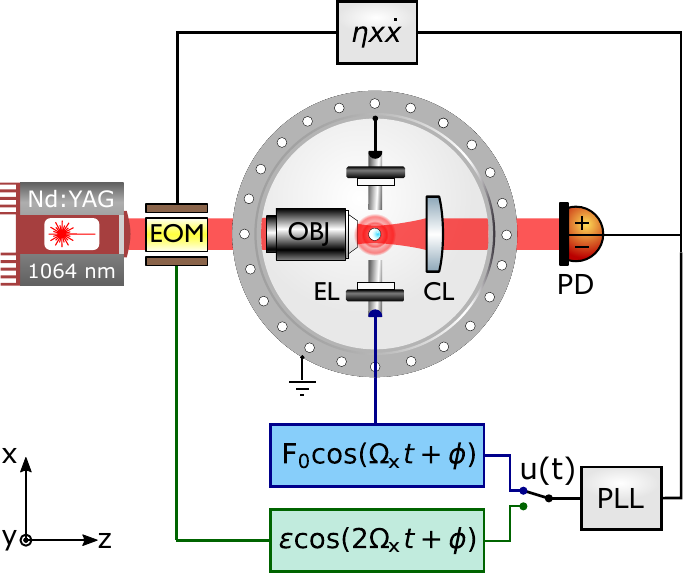}
	\caption{Experimental setup. A microscope objective (OBJ) with $\text{NA} = 0.8$ focuses a $\lambda=1064$~nm laser to trap a charged silica nanoparticle inside a vacuum chamber at $p=10^{-6}~\text{mbar}$. The particle position $x$ is measured by collecting (CL) and detecting the scattered light with a balanced photodiode (PD) \cite{gieseler_thesis}. The nanoparticle motion is stabilized through PFB ($\eta x\dot{x}$) by modulating the trap stiffness with an EOM. The same EOM drives the particle parametrically with $\varepsilon_m=\varepsilon\cos(2\Omega_xt)$. A pair of electrodes (EL) generate an oscillating Coulomb force of amplitude $F_0$ to directly drive the particle. We apply phase control by phase locking the output of the PLL $u(t)$ to the particle's motion signal $x(t)$ at a controlled phase difference $\phi$.  The PLL frequency equals the particle frequency $\Omega_x$ for direct driving (blue), and $2\Omega_x$ for parametric driving (green), respectively.}
	\label{fig:fig1_experimental_setup}
\end{figure}

For levitation, we use a strongly focused laser beam (wavelength $\lambda= 1064$ nm, optical power $P\simeq 100$ mW and numerical aperture $\text{NA}=0.8$) to trap a charged silica nanosphere of radius $R\approx 70$~nm at pressures $p=10^{-6}$ mbar (see Fig.~\ref{fig:fig1_experimental_setup}). The particle's charge $q_t=n_q\times q_e$ is controlled at single elementary charge precision $q_e=1.6\times 10^{-19}~\text{C}$ by ionizing gas molecules inside the vacuum chamber at moderate vacuum \cite{frimmer_controlling_2017}, where $n_q$ is the number of charges. The particle's centre-of-mass motion (c.o.m) is detected in balanced split detection \cite{gieseler_thesis} by collecting the scattered light with a collection lens (CL). The particle's motional eigenfrequencies are $(\Omega_x,\Omega_y,\Omega_z)/2\pi\simeq(125,140,40)~\text{kHz}$.
Parametric feedback (PFB) \cite{gieseler_subkelvin_2012, jain_direct_2016} provides nonlinear damping $\propto \eta x^2\dot{x}$, which stabilizes the particle's motion at low pressures and keeps the oscillation amplitudes in the linear regime.
In the following, we use independently direct and parametric excitation to drive the nanoparticle's motion into the nonlinear regime  along a single axis.
Along this direction the particle explores the trapping potential's anharmonicities \cite{gieseler_nonlin_mode_coupling_2014, gieseler_thermal_2013}, which are modeled by an additional cubic term in the restoring force $\propto \xi x^3$  \cite{Kovacic_Duffing}.

Hence, the dimensional equation of motion  
\begin{equation}\label{eq:eq_of_motion} 
\ddot{x}+(\Gamma_x+\Omega_x\eta x^2)\dot{x}+\Omega_x^2\left[1+\xi x^2+\varepsilon_{m}(t)\right]x=
\frac{F_{el}(t)}{m}
\end{equation}
describes the dynamics of the particle position $x$ along the x-axis.
Here, $\Gamma_x$ is the mechanical damping rate due to the surrounding gas \cite{hebestreit_calibration_2018, gieseler_thesis}, $m$ the oscillator's mass \cite{ricci_accurate_2019}, $\varepsilon_{m}$ a parametric modulation and $F_{el}$ an external electrostatic force.  Note that the stochastic force arising from residual air molecules is much weaker in high vacuum than the externally applied driving forces and can therefore be neglected.

\section{Phase control: implementation and measurements}
\paragraph{Direct phase control.}

For direct driving we apply a near resonant AC-voltage at $\Omega_d\approx \Omega_x$ to a pair of electrodes (EL), which surround the trapping region (Fig.~\ref{fig:fig1_experimental_setup}, blue wire) and are separated by $d=1.4~\text{mm}$ \cite{ricci_accurate_2019}. This exerts a Coulomb force $F_{el}(t)=F_0\cos(\Omega_d t)$ on the particle, where $F_0=q_tE_0$ is the driving strength and $E_0$ is the electric field amplitude along the x-axis at the trapping site.
We use a phase locked loop (PLL) to generate a driving signal $u(t)$ with a well defined phase $\phi$ with respect to the particle oscillation $x(t)$. 
The  oscillator's amplitude $\bar{x}(\phi)$ and frequency $\Omega(\phi)$ as a function of the phase are given by  (see Supplementary Material for derivation). 
\begin{subequations}
\label{eqn:phase_response_physical_direct}
	\begin{align}\label{eqn:phase_response_physical_direct_ampl} 
		\bar{x}(\phi) &\approx
		\left[4\frac{F_0}{m\Omega_x^2\eta}\sin(\phi)\right]^{1/3}\\ \label{eqn:phase_response_physical_direct_freq}
		\Omega(\phi) &\approx \Omega_x\left[1-\frac{F_0}{2m\Omega_x^2 \bar{x}(\phi)}\cos(\phi) +\frac{3}{8}\xi \bar{x}(\phi)^2  \right]		
	\end{align}
\end{subequations}
Since only positive amplitudes are physical, the phase values are restricted to $\phi\in[0,\pi]$.
Interestingly, from Eq.~\eqref{eqn:phase_response_physical_direct} it follows that the amplitude and frequency are unique functions of the phase, while the phase and amplitude are multi-valued functions of the frequency \cite{MillerJ_PhaseControl}.
Therefore, active phase control allows to explore the entire solution space. This is in contrast to frequency control, where the drive frequency is set independently and the oscillator phase can only assume the values corresponding to stable solutions \cite{gieseler_nonlin_mode_coupling_2014, ricci_optically_2017}. 

\begin{figure}[h!]
	\centering
	\includegraphics[width=0.95\linewidth]{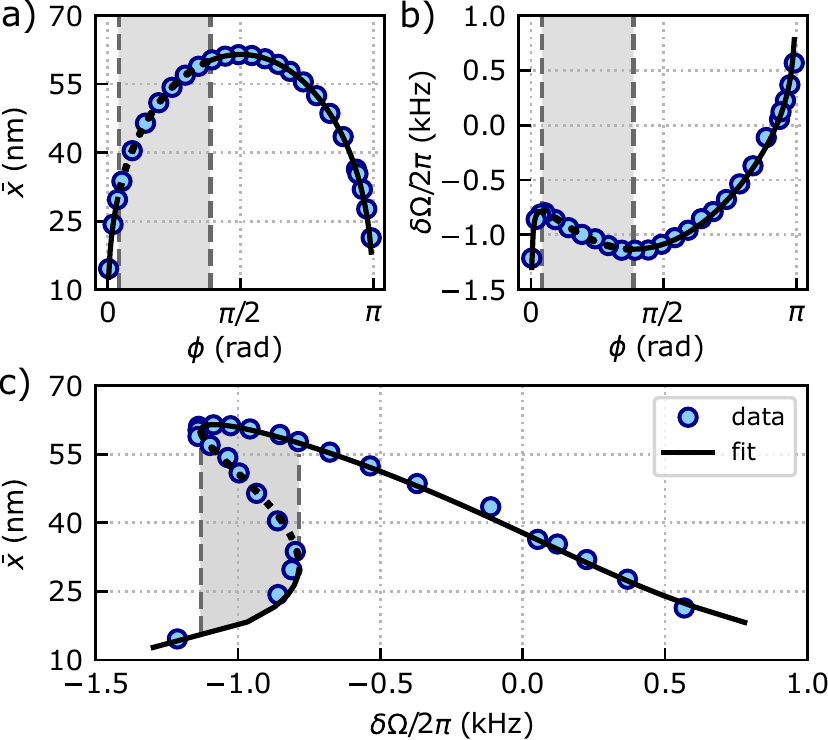}
	\caption{Direct phase control. a) Amplitude $\bar{x}(\phi)$ versus phase difference $\phi$. For small $\phi$ the oscillation amplitude $\bar{x}(\phi)$ increases with increasing $\phi$, reaching a maximum value of $\bar{x}(\phi)= 62~\text{nm}$ for $\phi= \pi/2$ before it drops towards the thermal amplitude for $\phi= \pi$. For $\phi>\pi$, we start cooling the nanoparticle, i.e. the amplitude drops below its thermal value (not shown here). For all $\phi$ there is only one amplitude solution. b) Frequency shift $\delta\Omega$ versus phase difference $\phi$. We observe a minimum $\delta\Omega/2\pi = -1.14~\text{kHz} $ at $\phi=0.38\pi$ which corresponds to the expected shift from Eq.~\eqref{eqn:phase_response_physical_direct_freq}. The part with the negative slope (gray shaded area) corresponds to the unstable branch. c) Amplitude response  $\bar{x}(\phi)$  versus frequency shift $\delta\Omega$. The bistable region (gray shaded area) has two stable (solid line) and one unstable branch (dashed line). The latter is only accessible with phase control, while the former leads to hysteresis in frequency control. The lines show the model fit to Eq.~\eqref{eqn:phase_response_physical_direct}.}
	\label{fig:direct_driving_main_text}
\end{figure}

In Fig.~\ref{fig:direct_driving_main_text}a we plot $\bar{x}(\phi)$ versus $\phi$ with a fit to Eq.~\eqref{eqn:phase_response_physical_direct_ampl}. The plot shows a maximum amplitude at $\phi = \pi/2$ with a unique amplitude solution for each phase value. The maximum amplitude can be tuned by changing the ratio $F_0/\eta$ for fixed trapping powers. In Fig.~\ref{fig:direct_driving_main_text}b we plot the frequency shift $\delta\Omega=\Omega(\phi)-\Omega_x$ versus the phase difference $\phi$ with a unique solution for all $\phi$. The region of negative slope between the local minimum and maximum defines the bistable regime in frequency control (gray shaded area). This regime has two stable (solid) and one unstable branch (dashed) as depicted in Fig.~\ref{fig:direct_driving_main_text}c, which lead to hysteresis in frequency control \cite{gieseler_nonlin_mode_coupling_2014}.
The width of the unstable region is given by $\delta\phi_{\text{us}} = \arctan{\left(\frac{\sqrt{1+\beta^2}-2}{\sqrt{\beta^2-3}}\right)}$, where $\beta = (3\xi/\eta)$.
Thus, the width of the unstable region remains constant for fixed $\beta$. However, the maximum frequency shift increases with $F_0$ (see Fig.~4a in the Supplementary Material). Thus, we can tune the maximum frequency shift without changing the width of the unstable region. This has applications in shaping the effective potential of the particle amplitude \cite{ricci_optically_2017}.\\

\paragraph{Parametric phase control.}
We complement our study with parametric excitation of the particle motion. An electro-optic modulator (EOM) (see Fig. 1) modulates the trap intensity, thereby modulating the trap stiffness  as $\varepsilon_{m}(t)=\varepsilon\cos(\Omega_p t)$ with $\Omega_p\approx 2\Omega_x$ and modulation depth  $\varepsilon$.
The solutions for $\bar{x}(\phi)$ and $\Omega(\phi)$ for the parametric drive are given by (see Supplementary Material for derivation):
\begin{subequations}
\label{eqn:phase_response_physical_parametric}
	\begin{align}
	\label{eqn:phase_response_physical_parametric_ampl}
		\bar{x}(\phi) & \approx \left[-2\frac{\varepsilon}{\eta}\sin(2\phi)\right]^{1/2} \\
		\label{eqn:phase_response_physical_parametric_freq}
		\Omega(\phi) &\approx\Omega_x\left[1+\frac{\varepsilon}{4}\sqrt{1+\beta^2}\cos\left(2\phi-\arctan \beta\right) \right]
	\end{align}
\end{subequations}
Similarly to direct driving, the oscillator's amplitude $\bar{x}(\phi)$ and frequency $\Omega(\phi)$ are unique functions of $\phi$.
Since only positive amplitudes are physical, the phase values are restricted to $\phi\in[-\pi/2,0]$.
For $\phi >0 $, the particle is no longer excited but cooled parametrically instead \cite{gieseler_21, jain_direct_2016}.
\begin{figure}[h!]
	\centering
	\includegraphics[width=0.85\linewidth]{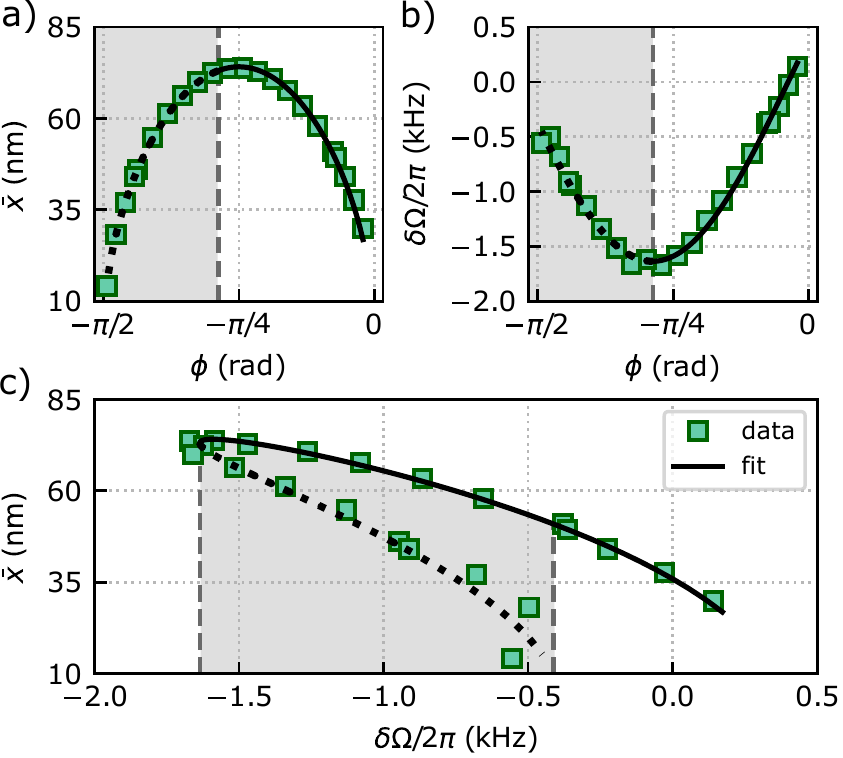}
	\caption{Parametric phase control. a) Amplitude $\bar{x}(\phi)$ versus phase difference $\phi$. $\bar{x}(\phi)$ increases with $\phi$, reaching a maximum value of $\bar{x}(\phi)\approx 72~\text{nm}$ at $\phi=-\pi/4$. For larger $\phi$ the amplitude declines towards its thermal value at $\phi=0$ and for $0<\phi<\pi/2$ the particle is cooled (not shown here). For all $\phi$ there is only one solution. b) Frequency shift $\delta\Omega$ versus $\phi$. The part with the negative slope (gray shaded area) corresponds to the unstable branch and the maximum frequency shift $\delta\Omega=-1.65~\text{kHz}$ is located at $\phi=\arctan{\beta}/2=-0.55\pi/2$. c) Amplitude response  $\bar{x}(\phi)$  versus $\delta\Omega$. The bistable region has one stable (solid line) amplitude and one unstable (dashed line) branch. The second stable branch corresponds to $\bar{x}=0$ and is not accessible in phase control. In frequency control the amplitude jumps between its thermal value and the stable branch when the drive frequency lies within the shaded area. The gray shaded area highlights the bistable region and the lines show the model fit to Eq.~\eqref{eqn:phase_response_physical_parametric}.}
	\label{fig:parametric_driving_main_text}
\end{figure}

\begin{figure}
	\centering
	\includegraphics[width=0.90\linewidth]{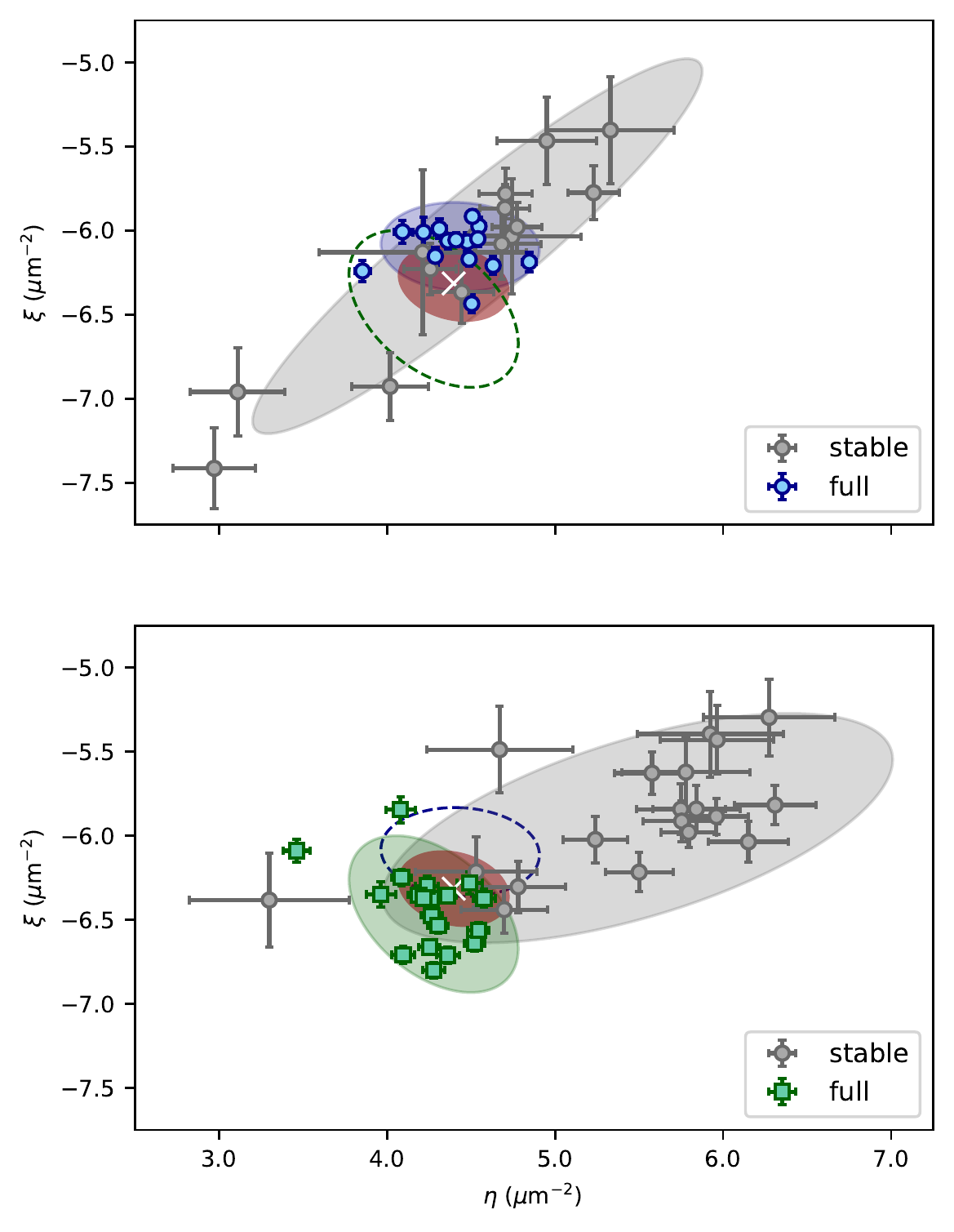}
	\caption{Estimates of the nonlinear parameters for a) direct (blue circles) and b) parametric driving (green squares) using the full nonlinear response and only the subset corresponding to the stable branches (gray circles) for various driving strengths, $F_0$ and $\varepsilon$, respectively. Each datapoint corresponds to a different driving strength and is obtained from a fit as the ones shown in Figs. \ref{fig:direct_driving_main_text} and \ref{fig:parametric_driving_main_text}. The shaded areas represent the mean and 1-sigma error ellipses of the datasets. The dashed green (blue) ellipse corresponds to the shaded green (blue) area in the other subplot. The white cross and red shaded area represent the mean and 1-sigma error ellipse of the combined parameter estimate, respectively. For the numerical values see Tab. \ref{tab:parameters}
	}\label{fig:method_comparison_main_text}
\end{figure}

Fig.~\ref{fig:parametric_driving_main_text} shows the data and fits to Eq.~\eqref{eqn:phase_response_physical_parametric}.
The maximum of $\bar{x}(\phi)$ is reached at $\phi=\pi/4$.
In Fig.~\ref{fig:parametric_driving_main_text}b we plot $\delta\Omega$ versus $\phi$. 
The maximum frequency shift $\delta\Omega$ is given by  $\frac{\varepsilon}{4}\sqrt{1+\beta^2}$ when $\phi=\arctan(\beta)/2$, yielding $\delta\Omega_{max} = -1.65~\text{kHz}$ for the data shown here.
In contrast to direct driving, $\delta\Omega$ has no local maximum and only a local minimum. The local minimum separates the stable from the unstable regime \cite{ricci_optically_2017, gieseler_nonlin_mode_coupling_2014}.  The $\bar{x}(\phi)$ versus $\delta\Omega(\phi)$ representation shows the characteristic bistability (Fig.~\ref{fig:parametric_driving_main_text}c), which we again highlight in gray. In contrast to the direct driving case, we only observe one stable branch because the second stable solution in parametric driving is $\bar{x}=0$ \cite{gieseler_nonlinear_2014}, which can not be accessed in phase control since there is no signal for the PLL to lock to.
\paragraph{Extracting nonlinear coefficients.} 
The nonlinear parameters $\xi$ and $\eta$ do not depend on the driving strength $F_0$, nor on the modulation depth $\varepsilon$. 
Thus, to obtain a richer dataset we repeat the measurements shown in Figs.\ref{fig:direct_driving_main_text} and \ref{fig:parametric_driving_main_text} for different driving strengths $F_0$ and modulation depths $\varepsilon$, respectively.
For each measurement, we extract the nonlinear parameters from a fit to our models Eq.~\eqref{eqn:phase_response_physical_direct} and Eq.~\eqref{eqn:phase_response_physical_parametric}.
The extracted values of the nonlinear parameters are shown in Fig~\ref{fig:method_comparison_main_text}.
The parameter estimates shown in blue (Eq.~\eqref{eqn:phase_response_physical_direct}) and green (Eq.~\eqref{eqn:phase_response_physical_parametric}) are obtained from a fit to the full dataset, while the gray values are obtained from a fit to a subset of the data that contains only the stable branches (solid lines in Figs.~\ref{fig:direct_driving_main_text} and \ref{fig:parametric_driving_main_text}).
The latter corresponds to the most complete dataset, that is available in methods without phase control, i.e. when the resonator is driven with a harmonic signal at fixed frequency \cite{gieseler_nonlin_mode_coupling_2014, ricci_optically_2017, eichler_parametric_2011, kozinsky_tuning_nodate} or excited thermally \cite{gieseler_thermal_2013, Fonseca_nonlin_paultrap, Latmiral_probingAnharmonicity_cavity_2016}.

Thermal excitations or other stochastic processes such as frequency fluctuations of the resonator, however, cause random transitions between the stable branches and quench the bistable region in frequency control. Thus, the observable data is reduced in practice if the stochastic forces are not kept at bay \cite{ricci_optically_2017}.

\begin{table*}[t!]
\centering
\arrayrulecolor{black}
\begin{tabular}{lcccc} 
\toprule
     & \multicolumn{2}{c}{full}&\multicolumn{2}{c}{stable}\\
     & $\eta$ [$\mu\text{m}^{-2}$] & $\xi$ [$\mu\text{m}^{-2}$] & $\eta $ [$\mu\text{m}^{-2}$]& $\xi$  [$\mu\text{m}^{-2}$] \\
direct&$4.394\pm0.010$&$-6.090\pm0.015$&$4.631\pm0.073$&$-6.254\pm0.064$\\
parametric&$4.240\pm0.016$&$-6.422\pm0.013$&$5.463\pm0.070$&$-5.882\pm0.040$\\
\bottomrule
combined&$\mathbf{4.317\pm0.009}$&$\mathbf{-6.256\pm0.010}$&$5.047\pm0.051$&$-6.068\pm0.038$\\
\bottomrule
\end{tabular}
\caption{Parameter estimates of the nonlinear coefficients. The estimates for direct and parametric driving, respectively, is the precision weighted mean of the individual model fits for each method. The combined values is the mean of the two methods weighted by the inverse of their covariance matrices. 
The first two columns correspond to the full dataset and give the centers of the green (direct), blue (parametric) and red (combined) ellipses in Fig. \ref{fig:method_comparison_main_text}. The errors equal the widths of the ellipses along the parameter axes. The values in the last two columns correspond to the subset of the data with only the stable branches and the gray ellipses in Fig. \ref{fig:method_comparison_main_text}.
\label{tab:parameters}}
\arrayrulecolor{black}
\end{table*}

The comparison of the full dataset and the subset allows us to quantify the improvement in parameter estimation that we get by making the unstable branch accessible through phase control. Clearly, the parameter estimates from using the full response function have much better precision (smaller error bars) and show a much smaller variance across the dataset.
The fact that the scatter of the datapoints is much larger than the statistical error from the fit suggests the presence of a systematic error when varying the driving strength. The covariance matrices of each dataset provide an estimate for the size of the systematic error and are shown as elliptical shaded area in Fig. \ref{fig:method_comparison_main_text}.
We also observe that the nonlinear parameters are strongly correlated. This makes it particularly challenging to get an accurate estimate of the nonlinear parameters in parametric driving when one observes only the stable branch.

For each method (direct and parametric drive), we calculate the precision weighted mean both for the full and for the reduced dataset. Then we combine the two results weighted by the inverse of their covariance matrices to obtain the best parameter estimate of the combined data $\eta=4.317\pm0.009 \:\mu\text{m}^{-2}$ and $\xi=-6.256\pm0.010\: \mu\text{m}^{-2}$, which are shown as white crosses in Fig. \ref{fig:method_comparison_main_text} together with the 1-sigma error ellipse (red).
The parameter estimates of the individual datasets are summarized in
Table \ref{tab:parameters}.



%
\section{Conclusion and outlook}
We have shown that phase control allows to access the full response function of a nonlinear oscillator and that the information contained in the unstable branch significantly improves the nonlinear parameter estimation.
We also demonstrated that parametric and direct driving give overlapping results within the measurement error. By combining the results from both methods we reduce the overall error and get a better estimate of the actual parameter values. 

The parameter estimation by phase control shown here is not restricted to levitated particles in vacuum but also applies to solid state systems such as carbon nanotubes \cite{eichler_nonlinear_2011} stressed, high-aspect ratio  \cite{catalini2021modeling} and other nanomechanical resonators \cite{shin2021spiderweb, GieselerMagnetoMag} where nonlinear effects are readily observable.
Levitated systems, however, will benefit from our sensitive technique for tracking and characterizing the parameter changes introduced by perturbations of the trapping potential near surfaces \cite{Rozenn_sublambda, winstone_electrostaticimage_force_sensing} or electrostatic  traps \cite{Fonseca_nonlin_paultrap}.
Moreover, our approach paves the way for new studies of the nonlinear regime, such as realizing a parametron with more than two states \cite{Frimmer_rapidflipping}.

\begin{acknowledgments}
M.T.C. and J.G. thank James M. L. Miller for stimulating discussions. The authors acknowledge financial support from the European Research Council through grant QnanoMECA (CoG-64790), Fundació Privada Cellex, CERCA Programme/Generalitat de Catalunya, and the Spanish Ministry of Economy and Competitiveness through the Severo Ochoa Programme for Centres of Excellence in R\&D (SEV-2015-0522).
\end{acknowledgments}


%
\clearpage

\end{document}